\documentclass[12pt]{article}
\usepackage{epsfig}
\def\be{\begin{equation}}
\def\ee{\end{equation}}
\def\bea{\begin{eqnarray}}
\def\eea{\end{eqnarray}}
\usepackage{graphicx}

\catcode`\@=11
\def\lsim{\mathrel{\mathpalette\@versim<}}
\def\gsim{\mathrel{\mathpalette\@versim>}}
\def\@versim#1#2{\vcenter{\offinterlineskip
\ialign{$\m@th#1\hfil##\hfil$\crcr#2\crcr\sim\crcr } }}
\catcode`\@=12

\catcode`@=12
\begin{document}
\thispagestyle{empty}
\begin{flushright}
UCRHEP-T543\\
August 2014\
\end{flushright}
\vspace{0.6in}
\begin{center}
{\LARGE \bf Scotogenic Inverse Seesaw Model\\ 
of Neutrino Mass\\}
\vspace{1.2in}
{\bf Sean Fraser, Ernest Ma, and Oleg Popov\\}
\vspace{0.2in}
{\sl Department of Physics and Astronomy, University of California,\\
Riverside, California 92521, USA\\}
\end{center}
\vspace{1.2in}
\begin{abstract}\
A variation of the original 2006 radiatve seesaw model of neutrino mass 
through dark matter is shown to realize the notion of inverse seesaw 
naturally.  The dark-matter candidate here is the lightest of three 
real singlet scalars which may also carry flavor.
\end{abstract}

\newpage
\baselineskip 24pt

In 1998, the simplest realizations of the dimension-five operator~\cite{w79} 
for Majorana neutrino mass, i.e. $(\nu_i \phi^0)(\nu_j \phi^0)$, were discussed 
systematically~\cite{m98} for the first time.  Not only was the nomenclature 
for the three and only three tree-level seesaw mechanisms established: 
(I) heavy singlet neutral Majorana fermion $N$~\cite{seesawI}, 
(II) heavy triplet Higgs scalar $(\xi^{++}, \xi^+, \xi^0)$~\cite{seesawII}, 
and (III) heavy triplet Majorana fermion 
$(\Sigma^+, \Sigma^0, \Sigma^-)$~\cite{seesawIII}, the three generic one-loop 
irreducible radiative mechanisms involving fermions and scalars were 
also written down for the first time.  Whereas one such radiative 
mechanism was already well-known since 1980, i.e. the Zee model~\cite{z80}, 
a second was not popularized until eight years later in 2006, when it was 
used~\cite{m06} to link neutrino mass with dark matter, called 
{\it scotogenic} from the Greek {\it scotos} meaning darkness.  The 
third remaining unused mechanism 
is the subject of this paper.  It will be shown how it is a natural 
framework for a scotogenic inverse seesaw model of neutrino mass, as 
shown in Fig.~1.
\begin{figure}[htb]
\vspace*{-3cm}
\hspace*{-3cm}
\includegraphics[scale=1.0]{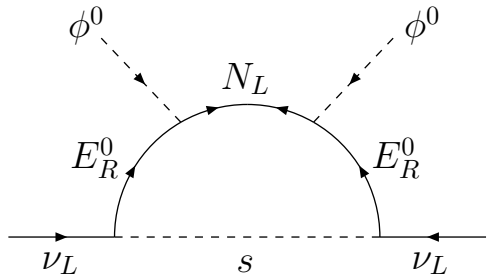}
\vspace*{-21.5cm}
\caption{One-loop generation of inverse seesaw neutrino mass.}
\end{figure}
The new particles are three real singlet scalars $s_{1,2,3}$, and one set of 
doublet fermions $(E^0, E^-)_{L,R}$, and one Majorana singlet fermion $N_L$, all 
of which are odd under an exactly conserved discrete symmetry $Z_2$.
This specific realization was designated T1-3-A with $\alpha = 0$ in 
the compilation of Ref.~\cite{rzy13}.  Note however that whereas 
$(E^0,E^-)_L$ is not needed to complete the loop, it serves the dual purpose 
of (1) rendering the theory to be anomaly-free and (2) allowing $E$ to 
have an invariant mass for the implementation of the inverse seesaw mechanism.

The notion of inverse seesaw~\cite{ww83,mv86,m87} is based on an extension 
of the $2 \times 2$ mass matrix of the canonical seesaw to a $3 \times 3$ 
mass matrix by the addition of a second singlet fermion.  In the space 
spanned by $(\nu, N, S)$, where $\nu$ is part of the usual lepton doublet 
$(\nu,l)$ and $N,S$ are singlets, all of which are considered left-handed, 
the most general $3 \times 3$ mass matrix is given by
\begin{equation}
{\cal M}_\nu = \pmatrix{0 & m_2 & 0 \cr m_2 & m_N & m_1 \cr 0 & m_1 & m_S}.
\end{equation}
The zero $\nu-S$ entry is justified because there is only one $\nu$ to 
which $N$ and $S$ may couple through the one Higgs field $\phi^0$. 
The linear combination which couples may then be redefined as $N$, and 
the orthogonal combination which does not couple is $S$.  If $m_{S,N}$ 
is assumed much less than $m_1$, then the induced neutrino mass is
\begin{equation}
m_\nu \simeq {m_2^2 m_S \over m_1^2}.
\end{equation}
This formula shows that a nonzero $m_\nu$ depends on a nonzero $m_S$, and 
a small $m_\nu$ is obtained by a combination of small $m_S$ and $m_2/m_1$. 
This is supported by the consideration of an approximate symmetry, i.e. 
lepton number $L$, under which $\nu, S \sim +1$ and $N \sim -1$. 
Thus $m_{1,2}$ conserve $L$, but $m_S$ breaks it softly by 2 units. 
Note that there is also a finite one-loop contribution from 
$m_N$~\cite{asy11,bdp12}.

Other assumptions about $m_1, m_S, m_N$ are also possible~\cite{m09}.  If 
$m_2, m_N << m_1^2/m_S$ and $m_1 << m_S$, then a double seesaw occurs 
with the same formula as that of the inverse seesaw, but of course with 
a different mass hierarchy.  If $m_1, m_2 << m_N$ and $m_1^2/m_N << m_S << 
m_1$, then a lopsided seesaw~\cite{m09} occurs with $m_\nu \simeq -m_2^2/m_N$ 
as in the canonical seesaw, but $\nu-S$ mixing may be significant, i.e. 
$m_1 m_2/m_S m_N$, whereas $\nu-N$ mixing is the same as in the canonical 
seesaw, i.e. $\sqrt{m_\nu/m_N}$.  In the inverse seesaw, $\nu-N$ mixing is 
even smaller, i.e. $m_\nu/m_2$, but $\nu-S$ mixing is much larger, i.e. 
$m_2/m_1$, which is only bounded at present by about 0.03~\cite{g13}.  
In the double seesaw, the effective mass of $N$ is $m_1^2/m_S$, 
so $\nu-N$ mixing is also $\sqrt{m_\nu/m_N}$.  Here $m_S >> m_N$, so the 
$\nu-S$ mixing is further suppressed by $m_1/m_S$.

In the original scotogenic model~\cite{m06}, neutrino mass is radiatively 
induced by heavy neutral Majorana singlet fermions $N_{1,2,3}$ as shown in 
Fig.~2.  
\begin{figure}[htb]
\vspace*{-3cm}
\hspace*{-3cm}
\includegraphics[scale=1.0]{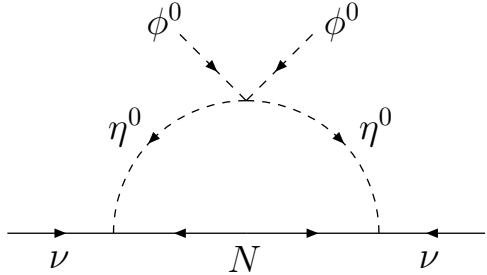}
\vspace*{-21.5cm}
\caption{One-loop generation of seesaw neutrino mass with heavy Majorana $N$.}
\end{figure}
However, they may be replaced by Dirac fermions.  In that case, a $U(1)_D$ 
symmetry may be defined~\cite{mpr13}, under which $\eta_{1,2}$ transform 
oppositely.  If $Z_2$ symmetry is retained, then a radiative inverse 
seesaw neutrino mass is also possible~\cite{hkmr07,ot12}. 
We discuss here instead the new mechanism of Fig.~1, based on the third 
one-loop realization of neutrino mass first presented in Ref.~\cite{m98}.
The smallness of $m_N$, i.e. the Majorana mass of $N_L$, may be naturally 
connected to the violation of lepton number by two units, as in the 
original inverse seesaw proposal using Eq.~(1).  It may also be a 
two-loop effect as first proposed in Ref.~\cite{m09-1}, with a number of 
subsequent papers by other authors, including Refs.~\cite{b11,lm12,ghl12}.

In our model, lepton number is carried by $(E^0, E^-)_{L,R}$ as well as $N_L$. 
This means that the Yukawa term $\bar{N}_L (E^0_R \phi^0 - E^-_R \phi^+)$ 
is allowed, but not $N_L (E^0_L \phi^0 - E^-_L \phi^+)$.  In the $3 \times 3$ 
mass matrix spanning $(\bar{E}^0_R, E^0_L, N_L)$, i.e.
\begin{equation}
{\cal M}_{E,N} = \pmatrix{0 & m_E & m_D \cr m_E & 0 & 0 \cr m_D & 0 & m_N},
\end{equation}
$m_E$ comes from the invariant mass term $(\bar{E}_R^0 E^0_L + E^+_R E^-_L)$, 
$m_D$ comes from the Yukawa term given above connecting $N_L$ 
with $E_R^0$ through $\langle \phi^0 \rangle = v$, and $m_N$ is the soft 
lepton-number breaking Majorana mass of $N_L$.  Assuming that $m_N << m_D, m_E$, 
the mass eigenvalues of ${\cal M}_{E,N}$ are
\begin{eqnarray}
m_1 &=& {m_E^2 m_N \over m_E^2 + m_D^2}, \\ 
m_2 &=& \sqrt{m_E^2 + m_D^2} + {m_D^2 m_N \over 2 (m_E^2 + m_D^2)}, \\ 
m_3 &=& -\sqrt{m_E^2 + m_D^2} + {m_D^2 m_N \over 2 (m_E^2 + m_D^2)}.
\end{eqnarray}
In the limit $m_N \to 0$, $E^0_R$ pairs up with $E_L^0 \cos \theta  + 
N_L \sin \theta$ to form a Dirac fermion of mass $\sqrt{m_E^2 + m_D^2}$, 
where $\sin \theta = m_D/\sqrt{m_E^2 + m_D^2}$.  This means that the 
one-loop integral of Fig.~1 is well approximated by
\begin{equation}
m_\nu = {f^2 m_D^2 m_N \over 16 \pi^2 (m_E^2 + m_D^2 - m_s^2)} 
\left[ 1 - {m_s^2 \ln ((m_E^2 + m_D^2)/m_s^2) \over (m_E^2 + m_D^2 - m_s^2)} 
\right].
\end{equation}
This expression is indeed of the form expected of the inverse seesaw.

The radiative mechanism of Fig.~1 is also suitable for supporting a 
discrete flavor symmetry, such as $Z_3$. 
Consider the choice
\begin{equation}
(\nu_i,l_i)_L \sim \underline{1}, \underline{1}', \underline{1}'', ~~~ 
s_{1} \sim \underline{1}, ~~~ (s_2 + i s_3)/\sqrt{2} \sim \underline{1}', 
~~~ (s_2 - i s_3)/\sqrt{2} \sim \underline{1}'', 
\end{equation}
with mass terms $m_s^2 s_1^2 + {m_s'}^2 (s_2^2 + s_3^2)$, then the induced 
$3 \times 3$ neutrino mass matrix is of the form
\begin{eqnarray}
{\cal M}_\nu &=& \pmatrix{f_e & 0 & 0 \cr 0 & f_\mu & 0 \cr 0 & 0 & f_\tau}  
\pmatrix{I(m_s^2) & 0 & 0 \cr 0 & 0 & I({m'_s}^2) \cr 0 & I({m'_s}^2) & 0} 
\pmatrix{f_e & 0 & 0 \cr 0 & f_\mu & 0 \cr 0 & 0 & f_\tau} \nonumber \\ 
&=& \pmatrix{f_e^2 I(m_s^2) & 0 & 0 \cr 0 & 0 & f_\mu f_\tau I({m'_s}^2) \cr 0 & 
f_\mu f_\tau I({m'_s}^2) & 0},
\end{eqnarray}
where $I$ is given by Eq.~(7) with $f^2$ removed. 
Let $l_{iR} \sim \underline{1}, \underline{1}', \underline{1}''$, 
then the charged-lepton mass matrix is diagonal using just the one 
Higgs doublet of the standard model, in keeping with the recent 
discovery~\cite{atlas12,cms12} of the 125 GeV particle.
To obtain a realistic neutrino mass matrix, we break $Z_3$ softly, i.e. 
with an arbitrary $3 \times 3$ mass-squared matrix spanning $s_{1,2,3}$,  
which leads to 
\begin{equation}
\pmatrix{1 & 0 & 0 \cr 0 & 1/\sqrt{2} & i/\sqrt{2} \cr 0 & 1/\sqrt{2} & 
-i/\sqrt{2}} O^T \pmatrix{I(m_{s1}^2) & 0 & 0 \cr 0 & I(m_{s2}^2) & 0 \cr 0 & 0 & 
I(m_{s3}^2)} O \pmatrix{1 & 0 & 0 \cr 0 & 1/\sqrt{2} & 1/\sqrt{2} \cr 0 & 
i/\sqrt{2} & -i/\sqrt{2}},
\end{equation}
where $O$ is an orthogonal matrix but not the identity, and there can be 
three different mass eigenvalues $m_{s1,s2,s3}$ for the $s_{1,2,3}$ sector.
The assumption of Eq.~(8) results in Eq.~(10) and allows the following 
interesting pattern for the neutrino mass matrix ${\cal M}_\nu$.  The 
Yukawa couplings $f_{e,\mu,\tau}$ may be rendered real by absorbing their 
phases into the arbitrary relative phases between $E_R^0$ and 
$\nu_{e,\mu,\tau}$.  If we further assume $f_2=f_3$, then ${\cal M}_\nu$ 
is of the form~\cite{m14}
\begin{equation}
{\cal M}_\nu = \pmatrix{A & C & C^* \cr C & D^* & B \cr C^* & B & D},
\end{equation}
where $A$ and $B$ are real.  Note that this pattern is protected by a 
symmetry first pointed out in Ref.~\cite{gl04}, i.e. $e \to e$ and 
$\mu - \tau$ exchange with $CP$ conjugation, and appeared previously in 
Refs.~\cite{m02,bmv03}.  As such, it is also guaranteed to yield maximal 
$\nu_\mu - \nu_\tau$ mixing $(\theta_{23} = \pi/4)$ and maximal $CP$ violation, 
i.e. $\exp (-i \delta) = \pm i$, whereas $\theta_{13}$ may be nonzero and 
arbitrary.  Our scheme is thus a natural framework for this possibility. 
Further, from Eq.~(7), it is clear that it is also a natural framework 
for quasi-degenerate neutrino masses as well.  Let  
\begin{equation}
F(x) = {1 \over 1-x} \left[ 1 + {x \ln x \over 1-x} \right],
\end{equation}
where $x = m_s^2/(m_E^2 + m_D^2)$, then Eq.~(7) becomes
\begin{equation}
m_\nu = {f^2 m_D^2 m_N \over (m_E^2 + m_D^2)} F(x).
\end{equation}
Since $F(0)=1$ and goes to zero only as $x \to \infty$, this scenario 
does not favor a massless neutrino.  If $f_{1,2,3}$ are all comparable 
in magnitude, the most likely outcome is three massive neutrinos with 
comparable masses.

Since the charged leptons also couple to $s_{1,2,3}$ through $E^-$, there 
is an unavoidable contribution to the muon anomalous magnetic moment 
given by~\cite{kt12}
\begin{equation}
\Delta a_\mu = {(g - 2)_\mu \over 2} = {f_\mu^2 m_\mu^2 \over 16 \pi^2 m_E^2} 
\sum_i |U_{\mu i}|^2 G(x_i),
\end{equation}
where
\begin{equation}
G(x) = {1 - 6x + 3x^2 + 2x^3 - 6x^2 \ln x \over 6 (1-x)^4},
\end{equation}
with $x_i = m_{si}^2/m_E^2$ and 
\begin{equation}
U = O \pmatrix{1 & 0 & 0 \cr 0 & 1/\sqrt{2} & 1/\sqrt{2} \cr 0 & i/\sqrt{2} 
& -i/\sqrt{2}}.
\end{equation}
To get an estimate of this contribution, let $x_i << 1$, then $\Delta a_\mu 
= f_\mu^2 m_\mu^2/96 \pi^2 m_E^2$.  For $m_E \sim 1$ TeV, this is of order 
$10^{-11} f_\mu^2$, which is far below the present experimental sensitivity 
of $10^{-9}$ and can be safely ignored.  The related amplitude for 
$\mu \to e \gamma$ is given by
\begin{equation}
A_{\mu e} = {e f_\mu f_e m_\mu \over 32 \pi^2 m_E^2} \sum_i U^*_{ei} U_{\mu i} 
G(x_i).
\end{equation}
Using the most recent $\mu \to e \gamma$ bound~\cite{meg13}
\begin{equation}
B = {12 \pi^2 |A_{\mu e}|^2 \over m_\mu^2 G_F^2} < 5.7 \times 10^{-13},
\end{equation}
and the approximation $\sum_i U^*_{ei} U_{\mu i} G(x_i) \sim 1/36$ (based on 
tribimaximal mixing with $x_1 \sim 0$ and $x_2 \sim 1$) and $m_E \sim 
1$ TeV, we find 
\begin{equation}
f_\mu f_e < 0.03.
\end{equation}
Let $f_{e, \mu, \tau} \sim 0.1$, $m_N \sim 10$ MeV, $m_D \sim 10$ GeV, 
$m_E \sim 1$ TeV, then the very reasonable scale of $m_\nu \sim 0.1$ eV 
in Eq.~(7) is obtained, justifying its inverse seesaw origin.

Since $N_L$ is the lightest particle with odd $Z_2$, it is a would-be 
dark-matter candidate.  However, suppose we add $N_R$ so that the two pair 
up to have a large invariant Dirac mass, then 
the lightest scalar (call it $S$) among $s_{1,2,3}$ is a dark-matter candidate.  
It interacts with the standard-model Higgs boson $h$ according to
\begin{equation}
-{\cal L}_{int} = {\lambda_{hS} \over 2} v h S^2 + {\lambda_{hS} \over 4} h^2 S^2.
\end{equation}
If we assume that all its other interactions are suppressed, then the 
annihilations $SS \to h \to$ SM particles and $SS \to hh$ determine  
its relic abundance, whereas its elastic scattering off nuclei via $h$ 
exchange determines its possible direct detection in underground experiments. 
A detailed analysis~\cite{cskw13} shows that the present limit of the 
invisible width of the observed 125 GeV particle (identified as $h$) 
allows $m_S$ to be only within several GeV below $m_h/2$ or greater than 
about 150 GeV using the recent LUX data~\cite{lux13}.  Note that the 
vector fermion doublet $(E^0,E^-)$ is not the usually considered vector 
lepton doublet because it is odd under $Z_2$ and cannot mix with the 
known leptons.

In conclusion, we have shown how neutrino mass and dark matter may be 
connected using a one-loop mechanism proposed in 1998.  This scotogenic 
model is naturally suited to implement the notion of inverse seesaw 
for neutrino mass, allowing the scale of new physics to be 1 TeV or 
less.  The imposition of a softly broken $Z_3$ flavor symmetry yields 
an interesting pattern of radiative neutrino mass, allowing for 
maximal $\theta_{23}$ and maximal $CP$ violation.  The real singlet 
scalars in the dark sector carry lepton flavor, the lightest of which is 
absolutely stable.  Our proposal provides thus a natural theoretical 
framework for this well-studied phenomenological possibilty.

This work is supported in part 
by the U.~S.~Department of Energy under Grant No.~DE-SC0008541.

\bibliographystyle{unsrt}

\end{document}